\documentclass[a4paper]{article}
\usepackage{amsfonts,amssymb,amsmath,amsthm,cite}
\usepackage{ucs} 
\usepackage[utf8x]{inputenc}
\usepackage{graphicx}
\usepackage[english]{babel}
\usepackage{slashed}
\usepackage{textcomp}
\usepackage{bm}

\evensidemargin=\oddsidemargin
\begin{document}
\begin{center}
	{\Large\textbf{On Two-Loop Effective Action of 2D Sigma Model}}
	\vspace{0.5cm}
	
	{\large P.~V.~Akacevich$^\ddag$~~~~~~A.~V.~Ivanov$^\dag$}
	
	\vspace{0.5cm}
	{\it St. Petersburg Department of Steklov Mathematical Institute of
		Russian Academy of Sciences,}\\{\it 27 Fontanka, St. Petersburg 191023, Russia}\\
	{\it Leonhard Euler International Mathematical Institute, 10 Pesochnaya nab.,}\\
	{\it St. Petersburg 197022, Russia}\\
	$\ddag${\it E-mail: pavel.akacevich@yandex.ru}\\
	$\dag${\it E-mail: regul1@mail.ru}
\end{center}
\vskip 10mm
\date{\vskip 20mm}

\begin{abstract}
In this paper, we study two-loop contribution to the effective action of a two-dimensional sigma model. We derive a new formula, which can be applicable to a regularization of general type. As examples, we obtain known results for dimensional regularization and investigate new types of cutoff one. Also, we discuss non-local contributions and restrictions on the regularization.
\end{abstract}

\normalsize
	
\tableofcontents

\section{Introduction}
One of the ways to study models in quantum field theory \cite{3,9,10} is the perturbative expansion of path integrals with respect to a small parameter. Unfortunately, such decompositions, as a rule, contain divergent integrals \cite{6,7,105}, which follow from an integration of generalized functions \cite{Gelfand-1964,Vladimirov-2002} and their nonlinear combinations. This fact leads to the necessity of introducing an appropriate regularization.

As practice shows, there are a lot of ways to introduce this. As examples, we can remember the dimensional regularization \cite{HV-72,19} and a cutoff one in the recent formulation \cite{Ivanov-Kharuk-2019,Ivanov-Kharuk-2020,Ivanov-Kharuk-2022,Iv-2022}, which appear in various calculations quite frequently. Of course, there are other different regularizations, which are not considered here and can be investigated separately, using our results. It should be noted that not every regularization keeps all the symmetries needed for a theory.

In the work, we study a two-loop contribution to the effective action of the standard two-dimensional sigma model \cite{sig2,sig1}. We derive a new formula for the divergent part, which can be used for an arbitrary regularization. Our paper generalizes the results of \cite{bagaev}, including some important corrections, in which two special regularizations were investigated. We make some additional essential remarks on a type of regularization in the model, discuss non-local contributions and non-logarithmic singularities.

The structure of the paper is the following. In Section \ref{sec:sigma}, we describe the basic concepts of the two-dimensional sigma model, introduce modified rules of the Feynman technique, derive the diagram representation for the two-loop effective action. Then, in Section \ref{sec:sigmares}, we study the infrared part of the two-loop contribution and derive a new formula. In Section \ref{sec:sigmaapp}, we study different popular types of infrared regularization and calculate the divergent parts using the new formula. Also, in the Remark we compare our result with one from \cite{bagaev}. In the last section we discuss the results and further steps.

\section{Sigma model}
\label{sec:sigma}
Let $\mathrm{SU}(n)$ be the special unitary group of degree $n\in\mathbb{N}$, see \cite{2}, and $\mathfrak{su}(n)$ is its Lie algebra.
Let $t^a$ be the generators of the algebra $\mathfrak{su}(n)$, where $a\in\{1,\ldots,\dim\mathfrak{su}(n)\}$,
such that the relations hold
\begin{equation}\label{constdef}
[t^a,t^b]=f^{abc}t^c,\,\,\,\,\,\,\mathrm{tr}(t^at^b)=-\frac{1}{2}\delta^{ab},
\end{equation}
where $f^{abc}$ are antisymmetric structure constants for $\mathfrak{su}(n)$ and '$\mathrm{tr}$' is the Killing form. We work with the adjoint representation. It is easy to verify that the structure constants have the following crucial properties
\begin{equation}\label{constprop}
f^{abc}f^{aef}=f^{abf}f^{aec}-f^{acf}f^{aeb},\,\,\,
f^{abc}f^{abe}=c_2\delta^{ce}.
\end{equation}

We work in the two-dimensional Euclidian space $\mathbb{R}^2$, the elements of which we notate by $x$ and $y$. At the same time, Greek letters $\mu,\nu$ denote the corresponding coordinate components. Further, we define classical and quantum actions of the sigma model \cite{sig1} using the following formulae
\begin{equation}\label{s1}
S[C]=\int_{\mathbb{R}^2}d^2x\,C_\mu^a(x)C_\mu^a(x)\,\,\,
\mbox{and}\,\,\,
W=-\ln\bigg(\int_{\mathcal{H}}\mathcal{D}g\,e^{-S[C]/4\gamma^2}\bigg),
\end{equation}
where $\gamma$ is a coupling constant, $C_{\mu}^a(x)t^a=(\partial_{x^\mu}^{\phantom{a}}g(x))g^{-1}(x)\in\mathfrak{su}(n)$ and
$g(x)\in\mathrm{SU}(n)$ for all $x\in\mathbb{R}^2$, and $\mathcal{H}$ is a functional space that can be defined using physical reasons.

Then, according to the main idea of the background field method \cite{102,103,24,25,26}, we make the following change of variable $g(x)=e^{\gamma\phi(x)}h(x)$ in the path integral (\ref{s1}). Also, for convenience, we introduce one more auxiliary gauge field $B_{\mu}^a(x)t^a=(\partial_{x^\mu}^{\phantom{a}}h(x))h^{-1}(x)\in\mathfrak{su}(n)$, its matrix-valued representation $B_{\mu}^{ab}(x)=f^{abc}B_{\mu}^b(x)$, the covariant derivative $D_{x^\mu}^{ab}=\partial_{x^\mu}^{\phantom{a}}\delta^{ab}-B_{\mu}^{ab}(x)$ corresponding to the gauge field, and the following Laplace-type operator $A^{ab}(x)=-D_{x_\mu}^{ab}\partial_{x^\mu}^{\phantom{a}}/2$.

Using the above, we can rewrite the classical action after the change of variable as
\begin{equation}\label{s2}
\frac{S[C]}{4\gamma^2}=\frac{S[B]}{4\gamma^2}+
\frac{1}{2\gamma}\Gamma_1[\phi]+\frac{1}{2}
\int_{\mathbb{R}^2}d^2x\,\phi^a(x)A^{ab}(x)\phi^b(x)
-\frac{\gamma}{12}\Gamma_3[\phi]-\frac{\gamma^2}{48}\Gamma_4[\phi]+\mathcal{O}(\gamma^3),
\end{equation}
where
\begin{equation}\label{s3}
\Gamma_1[\phi]=\int_{\mathbb{R}^2}d^2x\,
B_\mu^a(x)\partial_{x_\mu}^{\phantom{a}}\phi^a(x),\,\,\,
\Gamma_3[\phi]=-\int_{\mathbb{R}^2}d^2x\,f^{bea}
B_\mu^{bc}(x)\phi^c(x)\phi^e(x)\partial_{x_\mu}^{\phantom{a}}\phi^a(x),
\end{equation}
\begin{equation}\label{s4}
\Gamma_4[\phi]=\int_{\mathbb{R}^2}d^2x\,f^{abe}f^{cde}
\phi^a(x)\big(\partial_{x^\mu}^{\phantom{a}}\phi^b(x)\big)
\phi^c(x)D_{x_\mu}^{dh}\phi^h(x).
\end{equation}
Terms with $\gamma^n$ for $n\geqslant3$ are not used in the paper. They are presented by $S_n$, see the last formula in Section 2 of \cite{bagaev}, in which the minus before $1/2$ in the square brackets should be replaced by plus.

\begin{figure}[h]
	\centerline{\includegraphics[width=0.57\linewidth]{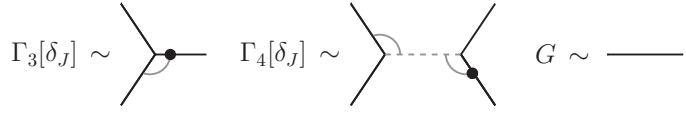}}
	\caption{Diagram technique elements.}
	\label{dte}
\end{figure}

Let us note that now we can introduce rules for the Feynman diagram technique. Using the fact that we are going to study only the two-loop contribution, we can define only three elements: vertices with three and four external lines, corresponding to the $\Gamma_3$ and $\Gamma_4$, and a line, corresponding to the Green's function. These elements are presented in Figure \ref{dte}. For convenience, the vertices are a little bit modified, see \cite{Ivanov-Kharuk-2022,13}, because they show the order of the indices in the structure constants, using the order of the lines, and the presence of the background field, using the dot.

Such type of change is quite useful, because it allows to apply the perturbative expansion to the quantum action. Indeed, let us substitute decomposition (\ref{s2}) into the path integral from (\ref{s1}) and calculate the standard Gaussian integrals, see \cite{Vas-98}. Then, using the fact that the gauge field $B_{\mu}(x)$ is the background field, which solves the quantum equation of motion and cancels all diagrams of glasses-type, we obtain the following expansion for the quantum action, or the \textit{effective action},
\begin{equation}\label{s5}
W[B]=\frac{S[B]}{4\gamma^2}+\frac{1}{2}\ln\det(A)-\gamma^2\bigg(\frac{1}{288}\Gamma_3^2\big[\delta_J\big]+
\frac{1}{48}\Gamma_4^{\phantom{2}}\big[\delta_J\big]\bigg)e^{g[J]}\bigg|_{J=0}^{\mbox{\scriptsize{1PI part}}}+
\mathcal{O}(\gamma^4).
\end{equation}
Here we have introduced one auxiliary functional as
\begin{equation}\label{s6}
g[J]=\frac{1}{2}\int_{\mathbb{R}^2}d^2x\int_{\mathbb{R}^2}d^2y\,J^a(x)G^{ab}(x,y)J^b(y),
\end{equation}
using the Green's function
\begin{equation}\label{s66}
	A^{ab}(x)G^{bc}(x,y)=\delta^{ac}\delta(x-y).
\end{equation}

\begin{figure}[h]
	\centerline{\includegraphics[width=0.75\linewidth]{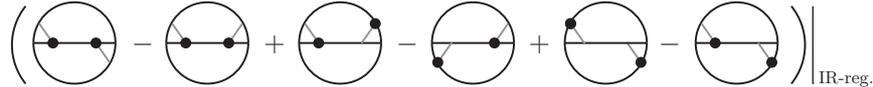}}
	\caption{Diagram representation of $W_{a}[B]$.}
	\label{wa}
\end{figure}

\begin{figure}[h]
	\centerline{\includegraphics[width=0.45\linewidth]{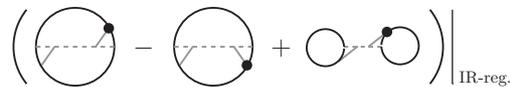}}
	\caption{Diagram representation of $W_{b}[B]$.}
	\label{wb}
\end{figure}

The second term in formula (\ref{s5}) is the one-loop correction. Its divergent part is very well known and can be written as $-c_2S[B]/16\pi\varepsilon$ in the case of the dimensional regularization, or with $L=\ln(\Lambda/\mu)$ instead of $1/\varepsilon$ for a cutoff one.
The third term corresponds to the two-loop contribution to the quantum action of the two-dimensional sigma model. Let us introduce some type of infrared regularization in the Green's function (in the region $x\sim y$) and two auxiliary objects
\begin{equation}
W_a[B]=\Gamma_3^2\big[\delta_J\big]e^{g[J]}\bigg|_{J=0}^{
	\substack{\mbox{\scriptsize{IR-reg.}}\\\mbox{\scriptsize{1PI part}}}}
,\,\,\,\,\,\,
W_b[B]=\Gamma_4^{\phantom{2}}\big[\delta_J\big]e^{g[J]}\bigg|_{J=0}^{
	\substack{\mbox{\scriptsize{IR-reg.}}\\\mbox{\scriptsize{1PI part}}}},
\end{equation}
which are depicted in Figures \ref{wa} and \ref{wb}.

It is quite interesting that the functionals $W_a$ and $W_b$ have the same diagram representation as in the case of the Yang--Mills theory. So, we can say that the two-dimensional sigma model is a lite version of the four-dimensional Yang--Mills theory, studied in papers \cite{3,Ivanov-Kharuk-2020,Ivanov-Kharuk-2022,13}. However, the difference between the two-loop contributions exists, because we need to study the linear combination of $W_a$ and $W_b$ with other coefficients. Hence, according to the main theory, we can define 
the regularized two-loop contribution $W_{2}^{\scriptsize{\mbox{reg}}}$ to the effective action using the following relation
\begin{equation}\label{s7}
W_{2}^{\scriptsize{\mbox{reg}}}[B]=-\frac{1}{288}W_a[B]
-\frac{1}{48}W_b[B].
\end{equation}

\section{Two-loop divergent part}
\label{sec:sigmares}
The main aim of this work is to derive a formula for the divergent part of the two-loop contribution, see (\ref{s7}), which can be applied to an arbitrary type of infrared regularization in the case, when the Seeley--DeWitt (sometimes they are named after Hadamard, Minakshisundaram \cite{111}, and Gilkey \cite{32}) coefficients \cite{1000,110} are not violated by the regularization.

To proceed we need to introduce some basic formulae for the Green's function expansion near the diagonal $x\sim y$. Also, for convenience, we will omit the notation of the group indices, meaning that we work with matrices. Then, in the two-dimensional case, we have the following expansion \cite{15,psi}
\begin{equation}\label{s8}
G(x,y)=2R_0(x-y)a_0(x,y)+2R_1(x-y)a_1(x,y)+2PS(x,y)+o(|x-y|^2),
\end{equation}
where $a_n(x,y)$ is the Seeley--DeWitt coefficient, which is constructed using the heat kernel method \cite{vas1,vas2} and can be found in \cite{30,31,32,33}.
Also, $\mu$ is an auxiliary parameter to make the combination $|x|\mu$ dimensionless,
\begin{equation}\label{s9}
R_0(x)=-\frac{1}{4\pi}\ln\big(|x|^2\mu^2\big),\,\,\,
R_1(x)=\frac{1}{16\pi}|x|^2\big(\ln\big(|x|^2\mu^2\big)-1\big),
\end{equation}
and $PS(x,y)$ is a smooth part. It is quite easy to verify that the last functions satisfy the following relations
\begin{equation}\label{s10}
-\partial_{x_\mu}\partial_{x^\mu}R_0(x)=\delta(x),\,\,\,
-\partial_{x_\mu}\partial_{x^\mu}R_1(x)=R_0(x)-\frac{1}{4\pi},\,\,\,
\partial_{x_\mu}R_1(x)=-\frac{x^\mu}{2}R_0(x),
\end{equation}
\begin{equation}\label{s11}
a_1(x,x)
=-\frac{1}{4}B_\mu(x)B_\mu(x)+\frac{1}{2}\partial_{x_\mu}B_\mu(x)
=-2A(x)a_0(x,y)\big|_{y=x}=8\pi A(x)PS(x,y)\big|_{y=x}.
\end{equation}
Moreover, we can claim that $a_0(x,y)$ is equal to a path-ordered exponential, see \cite{sig1}, which solves the following equation
\begin{equation}\label{s12}
(x-y)^\mu(\partial_{x^\mu}-B_\mu(x)/2)a_0(x,y)=0,\,\,\,a_0(x,x)=1.
\end{equation}

Now we can move on to the derivation of the main formula. Let us start with the contribution $W_a$. According to Figure \ref{wa} it consists of six diagrams. All these graphs have the same structure, so we comment on the calculation only for the first one. Other terms can be analyzed in the same way.

Let us note that to obtain the infrared singularity, we need to find a density with a negative power of $|x-y|$. It is possible only in the case, when either all Green's functions $G^{ab}(x,y)$ are replaced by $2\delta^{ab}R_0(x-y)$, or when two of them are replaced by $2\delta^{ab}R_0(x-y)$ and the third one is replaced by $2PS^{ab}(x,y)$, see formula (\ref{s8}). Hence, using the main idea of \cite{Ivanov-Kharuk-2019,Ivanov-Kharuk-2020,Ivanov-Kharuk-2022}, for the first diagram we get
\begin{equation*}\label{s13}
\int_{\mathbb{R}^d}d^dx\int_{\mathbb{R}^d}d^dy\,f^{abc}f^{deg}
\big(\partial_{x_\mu}G^{ag}(x,y)\big)
B_\mu^{bf}(x)G^{fh}(x,y)B_\nu^{eh}(y)
\partial_{y_\nu}G^{cd}(x,y)\Big|^{\scriptsize{\mbox{IR-reg.}}}
\stackrel{\mathrm{IR}}{=}\frac{8}{d}\mathrm{I}_1S[B]-\frac{8}{d}\mathrm{I}_2J_1[B],
\end{equation*}
where the sign $\stackrel{\mathrm{IR}}{=}$ means equality of singular parts, and
\begin{equation}\label{s14}
\mathrm{I}_1\stackrel{\mathrm{IR}}{=}c_2^2
\int_{\mathrm{B}_{1/\mu}}d^dx\,
\big(\partial_{x_\mu}R_0(x)\big)
\big(\partial_{x^\mu}R_0(x)\big)
R_0(x)
\Big|^{\scriptsize{\mbox{IR-reg.}}},
\end{equation}
\begin{equation}\label{s18}
\mathrm{I}_2\stackrel{\mathrm{IR}}{=}c_2
\int_{\mathrm{B}_{1/\mu}}d^dx\,
\big(\partial_{x_\mu}R_0(x)\big)
\big(\partial_{x^\mu}R_0(x)\big)
\Big|^{\scriptsize{\mbox{IR-reg.}}},
\end{equation}
\begin{equation}
\label{s15}
J_1[B]=
\int_{\mathbb{R}^d}d^dx\,B_\mu^{ab}(x)PS^{bc}(x,x)B_\mu^{ca}(x),
\end{equation}
\begin{equation}
\label{s19}
J_2[B]=\frac{1}{c_2}
\int_{\mathbb{R}^d}d^dx\,
f^{abc}f^{aed}
B_\mu^{bh}(x)B_\mu^{he}(x)PS^{cd}(x,x),
\end{equation}
\begin{equation}
\label{s20}
J_3[B]=
\int_{\mathbb{R}^d}d^dx\,PS^{aa}(x,x).
\end{equation}
Here, $\mathrm{B}_{1/\mu}$ is the ball at the origin with radius $1/\mu$. This is known feature of the two-loop calculation \cite{Ivanov-Kharuk-2019,Ivanov-Kharuk-2020,Ivanov-Kharuk-2022} that the radius can be selected equal to the fixed dimension parameter from (\ref{s9}). Besides that,
we have introduced some additional objects for further calculations. Also, we have changed the dimensional parameter $2$ by $d$ to be able to apply the dimensional regularization.
Then, repeating all calculations for the remaining diagrams, we get the following table of singularities, in which six columns correspond to the diagrams in Figure \ref{wa}.
\begin{center}
\begin{tabular}{c|rrrrrr}
$\mbox{Terms}$&$1$&$2$&$3$&$4$&$5$&$6$\\
\hline
$2\mathrm{I}_1S[B]/d$ & $4$ & $8$ & $-2$ & $2$ & $4$ & $2$\\
$2\mathrm{I}_2J_1[B]/d$ & $-4$ & $-4$ & $-2$ & $-2$ & $-4$ & $-2$\\
$2\mathrm{I}_2J_2[B]/d$ & $0$ & $-4$ & $4$ & $0$ & $0$ & $0$
\end{tabular}
\end{center}
Therefore, summing up all parts of the last table, we get the answer for the first two-loop part of the effective action
\begin{equation}
\label{s16}
W_a[B]\stackrel{\mathrm{IR}}{=}\frac{36}{d}\mathrm{I}_1S[B]-
\frac{36}{d}\mathrm{I}_2J_1[B].
\end{equation}
It is quite interesting to note that the non-local term $J_2[B]$ does not contribute to the divergent part. At the same time, the first non-local part $J_1[B]$ still remains. So, it should be cancelled by the contribution from $W_b[B]$ due to the locality of divergent terms of the model.

For further considerations we need to note some additional restrictions on the type of regularization. Using the fact that we work with the Green's functions on the diagonal, when $x\sim y$, it is necessary to require
\begin{equation}\label{s25}
R_0^{\phantom{2}}(x)\big|^{\scriptsize{\mbox{IR-reg.}}}_{x=0}
\,\,\,\mbox{and}\,\,\,
\partial_{x_\mu}\partial_{x^\mu}\Big(R_1^{\phantom{2}}(x)\big|^{\scriptsize{\mbox{IR-reg.}}}\Big)\Big|_{x=0}
\,\,\,\mbox{are finite}
\end{equation}
for all fixed values of the regularization parameter (not at the limit point). This restriction leads to the following additional properties
\begin{equation}\label{s26}
x^\mu\partial_{x^\mu}\Big(R_0^{\phantom{2}}(x)\big|^{\scriptsize{\mbox{IR-reg.}}}\Big)\Big|_{x=0}=
|x|^2\partial_{x_\mu}\partial_{x^\mu}\Big(R_0^{\phantom{2}}(x)\big|^{\scriptsize{\mbox{IR-reg.}}}\Big)\Big|_{x=0}=
0,
\end{equation}
\begin{equation}\label{s27}
\Big(\partial_{x_\mu}
\Big(R_0^{\phantom{2}}(x)\big|^{\scriptsize{\mbox{IR-reg.}}}\Big)\Big)\partial_{x^\mu}\Big(R_1^{\phantom{2}}(x)\big|^{\scriptsize{\mbox{IR-reg.}}}\Big)\Big|_{x=0}=0.
\end{equation}
Hence, we can omit the combinations mentioned above in further calculations.

Then, applying the manipulations to the second part $W_b[B]$, we get the following three contributions (columns), each of which relates to the corresponding diagram in Figure \ref{wb}. We draw attention that in the table we have missed the arguments and the abbreviation
$\big|^{\scriptsize{\mbox{IR-reg.}}}_{x=0}$ for simplicity. Also, we have introduced additional operator $N(x)=\partial_{x_\mu}\partial_{x^\mu}$.
\begin{center}
	\begin{tabular}{c|ccc}
		$\mbox{Terms}$&$1$&$2$&$3$\\
		\hline
		$c_2^2S[B]$ & $-R_0^2/2$ & $-R_0^2-R_0^{\phantom{2}}N R_1^{\phantom{2}}-
		R_1^{\phantom{2}}N R_0^{\phantom{2}}+
		R_0^{\phantom{2}}/4\pi$& $-R_0^2\phantom{\hat{T}}$ \\
		$c_2J_1[B]$ & $R_0^{\phantom{2}}$ & $0$& $2R_0^{\phantom{2}}$ \\
		$c_2J_2[B]$ & $0$ & $R_0^{\phantom{2}}+
		N R_1^{\phantom{2}}$& $0$ \\
		$4c_2J_3[B]$ & $0$ & $-NR_0^{\phantom{2}}$ & $0$
	\end{tabular}
\end{center}
Hence, after summing we get the result for the second term $W_b[B]$ in the form
\begin{align}\label{s17}
W_b[B]-W_b[0]\stackrel{\mathrm{IR}}{=}&\,
c_2^2S[B]\Big(-5R_0^2/2-R_0^{\phantom{2}}N R_1^{\phantom{2}}-
R_1^{\phantom{2}}N R_0^{\phantom{2}}+
R_0^{\phantom{2}}/4\pi\Big)(x)\Big|^{\scriptsize{\mbox{IR-reg.}}}_{x=0}\\\nonumber
&
+3c_2J_1[B]R_0^{\phantom{2}}(x)\Big|^{\scriptsize{\mbox{IR-reg.}}}_{x=0}
+c_2J_2[B]\Big(R_0^{\phantom{2}}+
N R_1^{\phantom{2}}\Big)(x)\Big|^{\scriptsize{\mbox{IR-reg.}}}_{x=0}
-4c_2J_3[B]N(x) R_0^{\phantom{2}}(x)\Big|^{\scriptsize{\mbox{IR-reg.}}}_{x=0}.
\end{align}
Therefore, with the use of formulae (\ref{s7}), (\ref{s16}), and (\ref{s17}), we can formulate the following final result:
\begin{align}\nonumber
W_2^{\mathrm{reg}}[B]-W_2^{\mathrm{reg}}[0]\stackrel{\mathrm{IR}}{=}&
-\frac{1}{8}S[B]\bigg(\frac{1}{d}\mathrm{I}_1-
\frac{c_2^2}{6}\Big(5R_0^2/2+R_0^{\phantom{2}}N R_1^{\phantom{2}}+
R_1^{\phantom{2}}N R_0^{\phantom{2}}-
R_0^{\phantom{2}}/4\pi\Big)(x)\Big|^{\scriptsize{\mbox{IR-reg.}}}_{x=0}\bigg)
\\\label{s22}&+
\frac{1}{8}J_1[B]\bigg(\frac{1}{d}\mathrm{I}_2-\frac{c_2}{2}R_0^{\phantom{2}}(x)\Big|^{\scriptsize{\mbox{IR-reg.}}}_{x=0}\bigg)
-\frac{c_2}{48}J_2[B]\Big(R_0^{\phantom{2}}+
N R_1^{\phantom{2}}\Big)(x)\Big|^{\scriptsize{\mbox{IR-reg.}}}_{x=0}
\\ \nonumber&
+\frac{c_2}{12}J_3[B]N(x) R_0^{\phantom{2}}(x)\Big|^{\scriptsize{\mbox{IR-reg.}}}_{x=0}.
\end{align}

Let us comment on some interesting properties of the last construction. Firstly, it has the non-logarithmic singularity, see the last term. Secondly, we have the presence of the both non-local parts, see $J_1[B]$ and $J_2[B]$. According to the main idea, the first one should cancel the corresponding term in $W_a[B]$, while the second one should not contribute itself.
All these issues we investigate in the next section, using some explicit regularizations and appropriate restrictions on them.

\section{Applications}
\label{sec:sigmaapp}

In this section we consider some examples, with the usage of the dimensional regularization and a cutoff one in the coordinate representation. To start the discussion we need to introduce two additional restrictions on the regularization
\begin{equation}\label{s24}
\mathrm{I}_2\stackrel{\mathrm{IR}}{=}c_2R_0^{\phantom{2}}(x)\Big|^{\scriptsize{\mbox{IR-reg.}}}_{x=0},\,\,\,\,\,\,
\Big(R_0^{\phantom{2}}+
N R_1^{\phantom{2}}\Big)(x)\Big|^{\scriptsize{\mbox{IR-reg.}}}_{x=0}
\stackrel{\mathrm{IR}}{=}0.
\end{equation}
Actually, the first one is quite natural condition, because it gives a connection between the deformations of the logarithm and its integral representation at the origin. The second condition leads to additional smoothness of $R_1$. These restrictions follow from formula (\ref{s22}) to cancel the non-local parts.

\subsection{Dimensional regularization}
Let us start with the most popular type of regularization, which appears in different multi-loop calculations quite frequently. According to the main idea, we need to deform our functions (\ref{s9}) with the use of an auxiliary dimensionless parameter $\varepsilon$.
\begin{align}\label{dr1}
R^{\varepsilon}_0(x)&=\frac{1}{4\pi}
\bigg(\frac{2\mu^{-\varepsilon}}{\varepsilon}+\pi^{\varepsilon/2}
\Gamma(-\varepsilon/2)|x|^{\varepsilon}\bigg)+\frac{\mu^{-\varepsilon}}{4\pi}(\gamma_0+\ln\pi),\\\label{dr2}
R^{\varepsilon}_1(x)&=\frac{|x|^2}{16\pi}
\bigg(\frac{-2\mu^{-\varepsilon}}{\varepsilon}+\pi^{\varepsilon/2}
\Gamma(-1-\varepsilon/2)|x|^{\varepsilon}\bigg)-
\frac{|x|^2\mu^{-\varepsilon}}{8\pi d}(\gamma_0+\ln\pi),
\end{align}
where $\gamma_0$ is the Euler--Mascheroni constant, $\varepsilon>0$ is small, and $d=2-\varepsilon$. It is quite easy to verify that the function $R^{\varepsilon}_i(x)$ tends to $R^{\phantom{\varepsilon}}_i(x)$ for $i=0,1$, when $\varepsilon\to+0$. Moreover, we can write out the following equalities
\begin{equation}\label{dr3}
-\partial_{x_\mu}\partial_{x^\mu}R^{\varepsilon}_0(x)=\delta(x),\,\,\,
-\partial_{x_\mu}\partial_{x^\mu}R^{\varepsilon}_1(x)=R^{\varepsilon}_0(x)-
\frac{\mu^{-\varepsilon}}{4\pi},
\end{equation}
which have the same form as ones in (\ref{s10}). From them we immediately receive
\begin{equation}\label{dr4}
\mathrm{I}_1\stackrel{\mathrm{IR}}{=}\frac{c_2^2}{2}\int d^dx\,\Big(-\partial_{x_\mu}\partial_{x^\mu}R^{\varepsilon}_0(x)
\Big)R^{\varepsilon}_0(x)R^{\varepsilon}_0(x)\stackrel{\mathrm{IR}}{=}
\frac{c_2^2}{2}\Big(R^{\varepsilon}_0(0)\Big)^2,
\end{equation}
\begin{equation}\label{dr5}
	\mathrm{I}_2\stackrel{\mathrm{IR}}{=}c_2\int d^dx\,\Big(-\partial_{x_\mu}\partial_{x^\mu}R^{\varepsilon}_0(x)
	\Big)R^{\varepsilon}_0(x)\stackrel{\mathrm{IR}}{=}c_2R^{\varepsilon}_0(0),
\end{equation}
where we have used the integration by parts several times.
Hence, we obtain the following answer
\begin{equation}\label{dr6}
\mbox{Pole of}\,\,
\big(W_{2}^{\scriptsize{\mbox{reg}}}[B]-W_{2}^{\scriptsize{\mbox{reg}}}[0]\big)
\stackrel{\mathrm{IR}}{=}-\frac{\varepsilon c_2^2}{64}S[B]\Big(R^{\varepsilon}_0(0)\Big)^2
\stackrel{\mathrm{IR}}{=}-
\frac{1}{\varepsilon}
\frac{c_2^2\mu^{-2\varepsilon}}{256\pi^2}S[B].
\end{equation}
This formula is consistent with the known one. We draw the attention that we have written the words “pole of\,” to omit discussion of $J_3{B}$, which follows together with $\delta(0)$. It is known that in the dimensional regularization such terms can be cancelled by an appropriate shift.

\textbf{Remark:} Despite the fact that our answer is consistent with one from \cite{bagaev}, we need to emphasize that we have two differences in the calculation process. The first one is a minus sign in the definition of the effective action.  The second one is the result for the integrals $\mathrm{I}_1$ and $\mathrm{I}_2$, which, actually, leads to the second minus. Compare  (\ref{dr4}), (\ref{dr5}) and the formulae from ($\Pi$.5a) in \cite{bagaev}.

\subsection{Cutoff regularization}
This type of regularization is a more natural, because it keeps the dimension of the space ($d=2$), although, at the same time, it violates the internal symmetries. Note that all calculations are produced in the coordinate representation. We want to test two the most popular cutoff regularizations, applicable in the presence of the local heat kernel (Seeley--DeWitt) coefficients.\\

\textbf{Cutoff-1.} This type appears in a number of recent works, see \cite{Ivanov-Kharuk-2019,Ivanov-Kharuk-2020,Ivanov-Kharuk-2022,Iv-2022, psi}, and shows very good results in comparison with known calculations. This is partially achieved due to the presence of free parameters in the regularization that can be used at the right time.

According to the main idea, we need to introduce the basic deformation $R_0^{\Lambda,1}$ for (\ref{s9}) with the usage of an auxiliary parameter
\begin{equation}\label{1:cala:10}
R_0^{\Lambda,1}(x)=\frac{\mathbf{f}\big(|x|^2\Lambda^2\big)}{4\pi}+\frac{1}{4\pi}
	\begin{cases}
		-\ln(|x|^2\mu^2),&|x|>1/\Lambda;\\
		\,\,\,\,\,\,\,\,\,\,\,\,\,\,
		2L,&|x|\leqslant1/\Lambda,
	\end{cases}
\end{equation}
where $\mathrm{supp}(\mathbf{f})\subset[0,1]$, $L=\ln(\Lambda/\mu)$, and the limit $\Lambda\to+\infty$ cancels the regularization. Of course, we need to require some additional properties for the function $\mathbf{f}$, because we want to have an appropriate smoothness, absence of the non-logarithmic part, and the combination $N(x)\mathbf{f}\big(|x-y|^2\Lambda^2\big)$ should tend to zero in the limit $\Lambda\to+\infty$ in the sense of generalized functions. So, we can write
the following set of conditions:
\begin{equation}\label{s33}
\mathbf{f}(\cdot)\in C^2\big([0,+\infty)\big),\,\,\,\mathbf{f}'(0)=0.
\end{equation}
We draw the attention, that the equalities $\mathbf{f}(1)=\mathbf{f}'(1)=0$ follow from the properties mentioned above, because the support of the function is in $[0,1]$.

Then, to keep generality, we do not introduce an explicit form for the deformation of $R_1^{\Lambda,1}$, with the exception of the second property from (\ref{s24}), which can be presented as $R_0^{\Lambda,1}(0)+NR_1^{\Lambda,1}(0)\to\alpha/4\pi\in\mathbb{R}$, when $\Lambda\to+\infty$. Further, using the formulae mentioned above we get the following answers
\begin{equation}\label{s28}
\mathrm{I}_1\stackrel{\mathrm{IR}}{=}
\frac{c_2^2}{8\pi^2}\bigg(L^2+L\int_0^1ds\,s\Big(\mathbf{f}'(s)\Big)^2\bigg),\,\,\,
\mathrm{I}_2\stackrel{\mathrm{IR}}{=}\frac{c_2L}{2\pi},\,\,\,
NR_0^{\Lambda,1}(0)=0,
\end{equation}
and, finally, we obtain the two-loop divergent part in the form
\begin{equation}\label{s29}
\big(W_{2}^{\scriptsize{\mbox{reg}}}[B]-W_{2}^{\scriptsize{\mbox{reg}}}[0]\big)
\stackrel{\mathrm{IR}}{=}-
\frac{c_2^2LS[B]}{64\pi^2}
\hat{\alpha},
\end{equation}
where
\begin{equation}\label{s35}
\hat{\alpha}=
	\bigg(\frac{1-\alpha-3\mathbf{f}(0)}{6}+\frac{1}{2}\int_0^1ds\,s\Big(\mathbf{f}'(s)\Big)^2\bigg).
\end{equation}
The last formula contains a few parameters, which can be fixed using some additional requirements. Unfortunately, the two-loop part does not provide enough limitation, so we are interested in the three-loop calculations, which may include some additional relations. As an example, we can take the explicit cutoff in the coordinate representation \cite{Iv-2022}, which leads to $\mathbf{f}(\cdot)=0$, $\alpha=1$, and $\hat{\alpha}=0$.\\

\textbf{Cutoff-2.} The last example describes one more deformation $|\cdot|^2\to|\cdot|^2+1/\Lambda^2$, appeared in the context of the four-dimensional Yang--Mills theory \cite{1-1-1,Ivanov-Kharuk-2022}. In this case we have
\begin{equation}\label{s31}
	R_0^{\Lambda,2}(x)=-\frac{1}{4\pi}\ln\big(|x|^2\mu^2+\mu^2/\Lambda^2\big),\,\,\,
	R_1^{\Lambda,2}(x)=\frac{|x|^2+1/\Lambda^2}{16\pi}\big(\ln\big(|x|^2\mu^2+\mu^2/\Lambda^2\big)-1\big).
\end{equation}
Then, making a set of simple calculations, we get the following answers
\begin{equation}\label{s30}
NR_0^{\Lambda,2}(0)=-\frac{\Lambda^2}{\pi},\,\,\,
NR_1^{\Lambda,2}(0)=-R_0^{\Lambda,2}(0)=-\frac{L}{2\pi},
\end{equation}
\begin{equation}\label{s34}
\mathrm{I}_1\stackrel{\mathrm{IR}}{=}-\frac{c_2^2}{16\pi^2}\int_{1/\Lambda^2}^{1/\mu^2}ds\,\frac{\ln\big(s\mu^2\big)}{s}\big(1-1/s\Lambda^2\big)\stackrel{\mathrm{IR}}{=}\frac{c_2^2}{8\pi^2}\big(L^2-L\big),\,\,\,
\mathrm{I}_2\stackrel{\mathrm{IR}}{=}c_2R_0^{\Lambda,2}(0)=\frac{c_2L}{2\pi},
\end{equation}
\begin{equation}\label{s32}
	\big(W_{2}^{\scriptsize{\mbox{reg}}}[B]-W_{2}^{\scriptsize{\mbox{reg}}}[0]\big)
	\stackrel{\mathrm{IR}}{=}
	\frac{c_2^2LS[B]}{128\pi^2}-\frac{c_2\Lambda^2}{12\pi}J_3[B].
\end{equation}
We see that the final answer includes non-logarithmic divergence. This fact is a negative feature of the mentioned regularization and restricts the usage.

In conclusion of this section, we want to summarize our results for all three regularizations. We have emphasized four categories: value of the main pole, absence/presence of logarithmic non-local part, non-logarithmic part, and free parameters.
\begin{center}
	\begin{tabular}{c|ccc}
		Feature&\footnotesize{\textbf{Dimensional}}&\footnotesize{\textbf{Cutoff-1}}&\footnotesize{\textbf{Cutoff-2}}\\
		\hline
		Pole$\,\,=-c_2^2S[B]/64\pi^2\times$ & $\mu^{-2\varepsilon}/4\varepsilon$ & $L\hat{\alpha}$& $-L/2\phantom{\hat{T}}$ \\
		\footnotesize{Absence of logarithmic non-local part} & yes &yes& yes \\
		\footnotesize{Absence of non-logarithmic part} & no & yes& no \\
		\footnotesize{Presence of free parameters} &no&yes&no
	\end{tabular}
\end{center}
Thus, different regularizations can give different results. Moreover, using free auxiliary parameters, introduced into the regularization procedure by hand, we can actually obtain various values of the main pole.

\section{Conclusion}
In this paper, we have studied the singular part of the two-loop effective action for the two-dimensional sigma model. We have obtained new formula (\ref{s22}), which can be used for an arbitrary type of the infrared (near the diagonal) regularization. In the last part of the work we have discussed some popular regularizations, dimensional and cut off one. We described their similarities and differences and compared the obtained results with the known. Moreover, we have noted some useful restrictions on the regularization, see (\ref{s25}) and (\ref{s24}), which appear quite naturally in this simple model.

Then, we want to comment on one interesting fact. We see that formulae (\ref{dr1}) and (\ref{dr2}) contain the Euler--Mascheroni constant, while the final answer for the divergent part does not. Actually, this follows from the invariance of the two-loop divergent part with respect to a special shift of the Green's function. Indeed, it is easy to verify that if we make the following shift with the use of the local special zero modes \cite{psi}
\begin{equation}\label{fun1}
G(x,y)\to G(x,y)+\rho\,\sum_{n=0}^{+\infty}
	\frac{(-1)^n|x-y|^{2n}}{4^n\Gamma(n+1)}a_n(x,y),
\end{equation}
where $\rho$ is an auxiliary fixed constant, the divergent part will have the same value. The same property was recently noticed in the four-dimensional Yang--Mills theory \cite{Kharuk-2021}. Actually, this feature follows from the fact that the two-loop divergent part is proportional to the logarithm $\ln(\Lambda/\mu)$. At the same time, the additional zero mode leads to a shift $\mu\to\mu_1$ in (\ref{s8}) and (\ref{s9}). Thus, using the property of the logarithm $\ln(\Lambda/\mu_1)=\ln(\Lambda/\mu)+\ln(\mu/\mu_1)$, we obtain the same singularity.

\vskip 5mm
\textbf{Acknowledgments.} Both authors are supported by the Ministry of Science and Higher Education of the Russian Federation, grant  075-15-2022-289. Also, A.V.Ivanov is supported in parts by the Foundation for the Advancement of Theoretical
Physics and Mathematics “BASIS”, grant “Young Russian Mathematics”. The authors are grateful to Natalia V. Kharuk for thoughtful reading of the manuscript, comments, and collaboration on related topics.

\end{document}